\documentclass{IEEEtran4Temp}
\usepackage{cite}
\usepackage{amssymb,amsfonts}
\usepackage{algorithmic}
\usepackage{graphicx}
\usepackage{textcomp}
\usepackage{xcolor}
\usepackage{subcaption}
\usepackage[cmex10]{amsmath}
\makeatletter
\let\old@ps@headings\ps@headings
\let\old@ps@IEEEtitlepagestyle\ps@IEEEtitlepagestyle
\def\psccfooter#1{%
    \def\ps@headings{%
        \old@ps@headings%
        \def\@oddfoot{\strut\hfill#1\hfill\strut}%
        \def\@evenfoot{\strut\hfill#1\hfill\strut}%
    }%
    \def\ps@IEEEtitlepagestyle{%
        \old@ps@IEEEtitlepagestyle%
        \def\@oddfoot{\strut\hfill#1\hfill\strut}%
        \def\@evenfoot{\strut\hfill#1\hfill\strut}%
    }%
    \ps@headings%
}
\makeatother

\begin{document}
%
\title{Oscillations between Grid-Forming Converters in Weakly Connected Offshore WPPs}
\author{
    \IEEEauthorblockN{Sulav Ghimire\textsuperscript{1,2,*}, Kanakesh V. Kkuni\textsuperscript{1}, Gabriel M. G. Guerreiro\textsuperscript{1,2},\\ Emerson D. Guest\textsuperscript{1}, Kim H. Jensen\textsuperscript{1}, Guangya Yang\textsuperscript{2}}
    \IEEEauthorblockA{\textsuperscript{1}\textit{Siemens Gamesa Renewable Energy A/S}, 7330 Brande, Denmark}
    \IEEEauthorblockA{\textsuperscript{2}\textit{Technical University of Denmark}, 2800 Kgs Lyngby, Denmark}
    \IEEEauthorblockA{\textsuperscript{*}Correspondence via: \textit{sulav.ghimire@siemensgamesa.com}}
}


\maketitle


\begin{abstract}
This paper studies control interactions between grid-forming (GFM) converters exhibited by power and frequency oscillations in a weakly connected offshore wind power plant (WPP). Two GFM controls are considered, namely virtual synchronous machine (VSM) and virtual admittance (VAdm) based GFM. The GFM control methods are implemented in wind turbine generators (WTGs) of a verified aggregated model of a WPP and the control interaction between these GFM WTGs is studied for several cases: cases with the same GFM control methods, and cases with different GFM control methods. A sensitivity analysis is performed for the observed oscillations to understand which system parameter affects the oscillations the most. Several solution methods are proposed and the inapplicability of some of the conventional solution methods are elaborated in this paper.
\end{abstract}



\section{Introduction}
\IEEEPARstart{T}{ype}-IV wind turbine generators (WTGs) consist of permanent magnet synchronous machine (PMSM) and full back-to-back converters. These WTGs have a complex control and dynamical structure and exhibit different control modes associated with the different controls and physical devices they have. When the natural frequencies of the different control modes match with each other, resonances and/or oscillations can be observed. Similarly, the generator/turbine side control modes and network converter side control modes can interact/resonate with each other to lead to torsional vibration modes and torsional oscillations \cite{9494824}. Further, the control modes of one WTG can also interact with that of another WTG leading to power and frequency oscillations. This paper studies the power and frequency oscillations between two WTGs and investigates the effect of network and grid parameters on these oscillations.

Conventional power systems exhibited low-frequency oscillations where multiple synchronous machines form a group and oscillate with or against another group of synchronous machines \cite{kundur2017power}. Similarly, low-frequency and sub-synchronous oscillations in type-IV WTGs have been a topic of discussion in the academic literature \cite{8742900}, and has also been observed in real-life power systems. For example, 4 Hz low-frequency oscillations were observed in wind power plants (WPPs) connected to the ERCOT grid \cite{6344713}, and 30 Hz sub-synchronous oscillations were observed in WPPs connected to Xinjiang Uygur Autonomous Region, China \cite{7878693}. Similar oscillations observed in grid-following (GFL) WPPs were found to be related to the PLL states and weak grid interconnections \cite{8742900, 7944679}. Type-IV wind turbines with GFM converter also exhibit torsional vibrations which can be mitigated via a damping filter and this solution can be reinforced by over-sizing the DC capacitor or by adding storage \cite{9494824}. Solutions to torsional vibration via a band pass filter and model-based damper are presented in \cite{6361459}. 


Systems such as WPPs with different GFM controls or neighboring WPPs with different GFM controls can also lead to control oscillations, although this hasn't been studied extensively in the academic literature or in the industry. Fast-acting controls of the WTG converters can interact with each other leading to power and frequency oscillations, eventually resulting in oscillatory instability. Avoiding such control interactions is crucial for the reliable operation of modern power systems. The low-frequency intra-turbine oscillation modes are usually operating point dependent and can be mitigated with DC-bus over-sizing \cite{9494824}, PSS, and damping washout filter \cite{9625987}. However, the inter-turbine oscillation modes investigated in this paper are near-synchronous and super-synchronous modes which are highly dependent on system topology rather than on operating points, and thus require a different approach. Since the low-frequency intra-turbine oscillation modes are operating point-dependent, they require extensive study over a wide range of operating points. However, the less dependency of these high-frequency oscillation modes on operating points gives an advantage in a number of studies required to solidify the findings.



Multiple original equipment manufacturers (OEMs,) vendors, and developers participate in modern power grids. In offshore WPPs, the controls approaches the different developers and OEMs employ in their GFM WPPs could also be different in the future. These GFM controls of the different WPPs can thus have control interactions with each other leading to different oscillatory behavior. In a system with multiple GFM converters, this paper provides an insight into how these control methods interact with each other in weak grids. This paper presents studies under several cases of control interaction: cases illustrating control interactions between identical GFM controls, cases with same GFM control with different control tunings, and cases with different GFM control architecture. Several methods to solve the arising oscillations are proposed in this paper.

The remainder of this paper is as follows. In section \ref{sec: System Description and Modelling}, a study framework is proposed where an aggregated model of an offshore WPP is chosen as a test network. A brief description of the system network topology and the GFM control methods used is also presented. Power and frequency oscillations between the WTGs with the same GFM controls are presented in Section \ref{sec: Oscillation between WTGs with same GFM controls}. Section \ref{sec: Oscillation between WTGs with different GFM controls} presents the oscillation between WTGs with different GFM controls where studies like fast Fourier transform (FFT) of the oscillations and a sensitivity analysis against the network parameters are performed. In Section \ref{sec: Potential Solution Methods}, potential solution methods for the mitigation of the presented oscillations are presented. Finally, the conclusions are drawn and future steps are noted in Section \ref{sec: Conclusion}.

\section{System Description and Modelling}\label{sec: System Description and Modelling}
\begin{figure}[htbp]
    \centering
    \includegraphics[width = 0.49\textwidth]{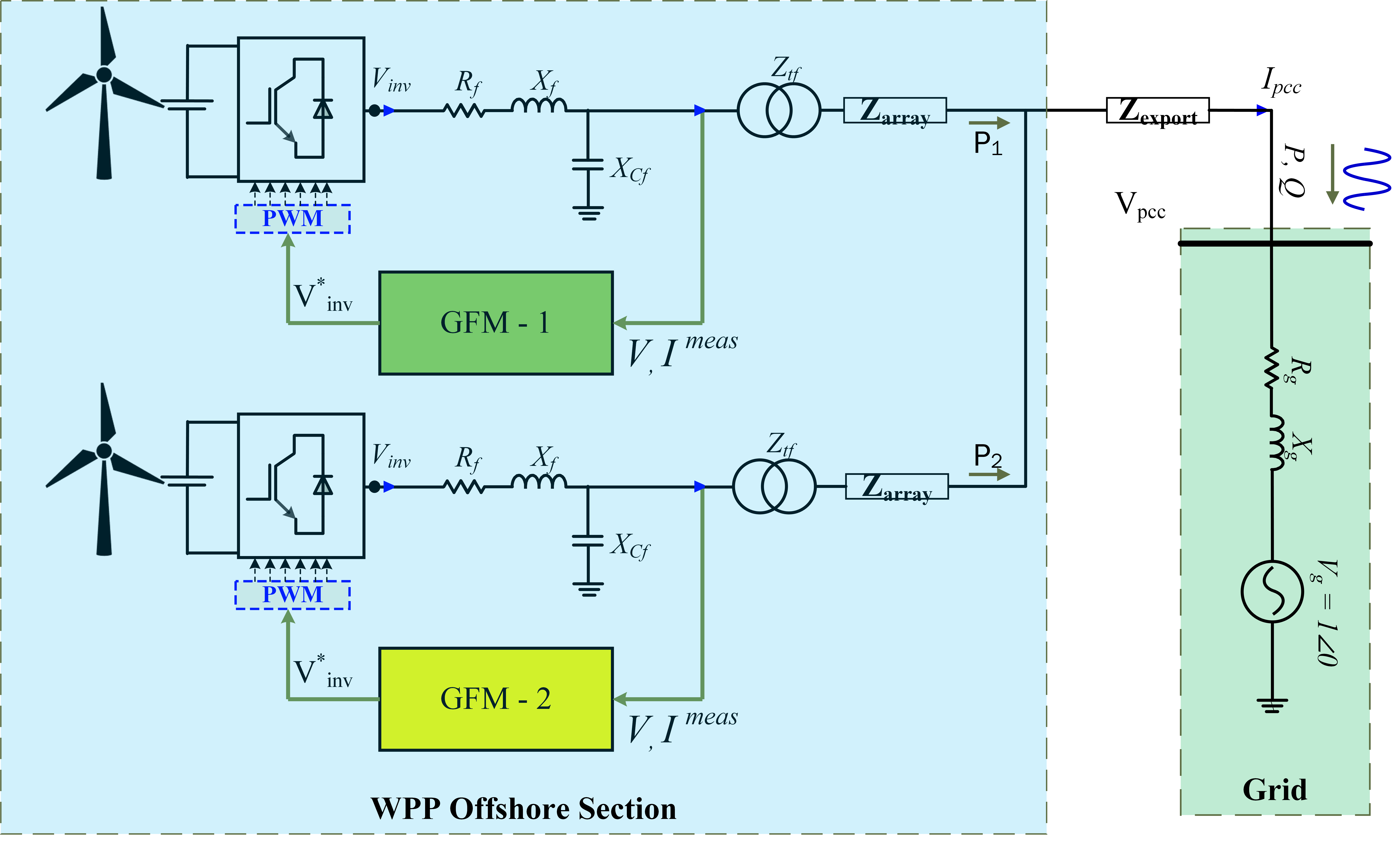}
    \caption{Network layout for GFM control interaction studies.}
    \label{fig: TwoConv}
\end{figure}

A network with two WTGs, each controlled by a GFM algorithm, connected to the grid with a nominal SCR value of 1.6 and $X_g/R_g=5$ (shown in Fig. \ref{fig: TwoConv}) is considered for this paper. The power system network used here is adapted from a WPP model with two power trains, each of which is connected to two aggregated models of WTG arrays with GFM controls. The control interactions between these WTGs are observed in terms of power and frequency oscillations and are studied further.

\begin{figure}[htbp]
    \centering
    \begin{subfigure}[b]{0.49\textwidth}
         \centering
         \includegraphics[width=\textwidth]{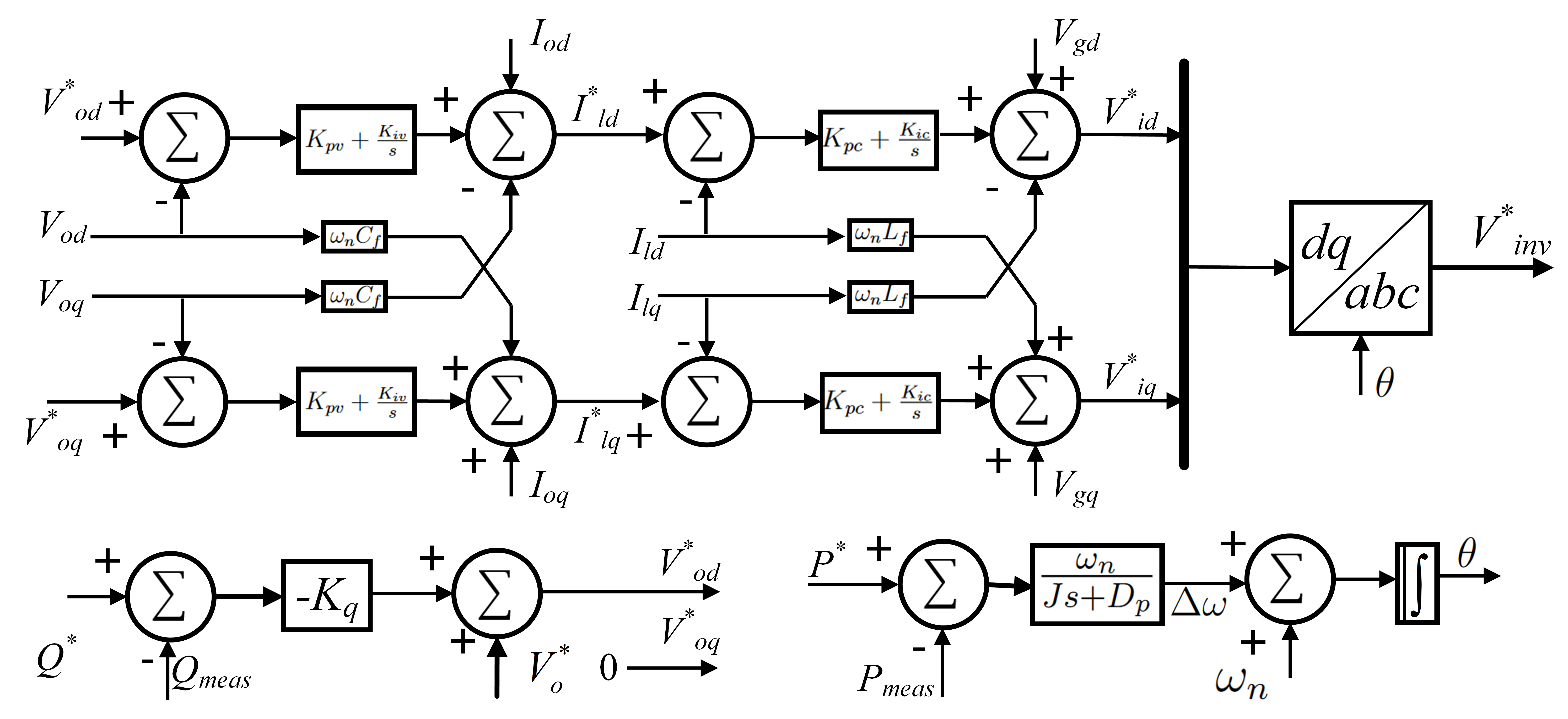}
         \caption{VSM control diagram}
         \label{fig: VSM Control Diagram}
     \end{subfigure}
     \hfill
     \begin{subfigure}[b]{0.49\textwidth}
         \centering
         \includegraphics[width=\textwidth]{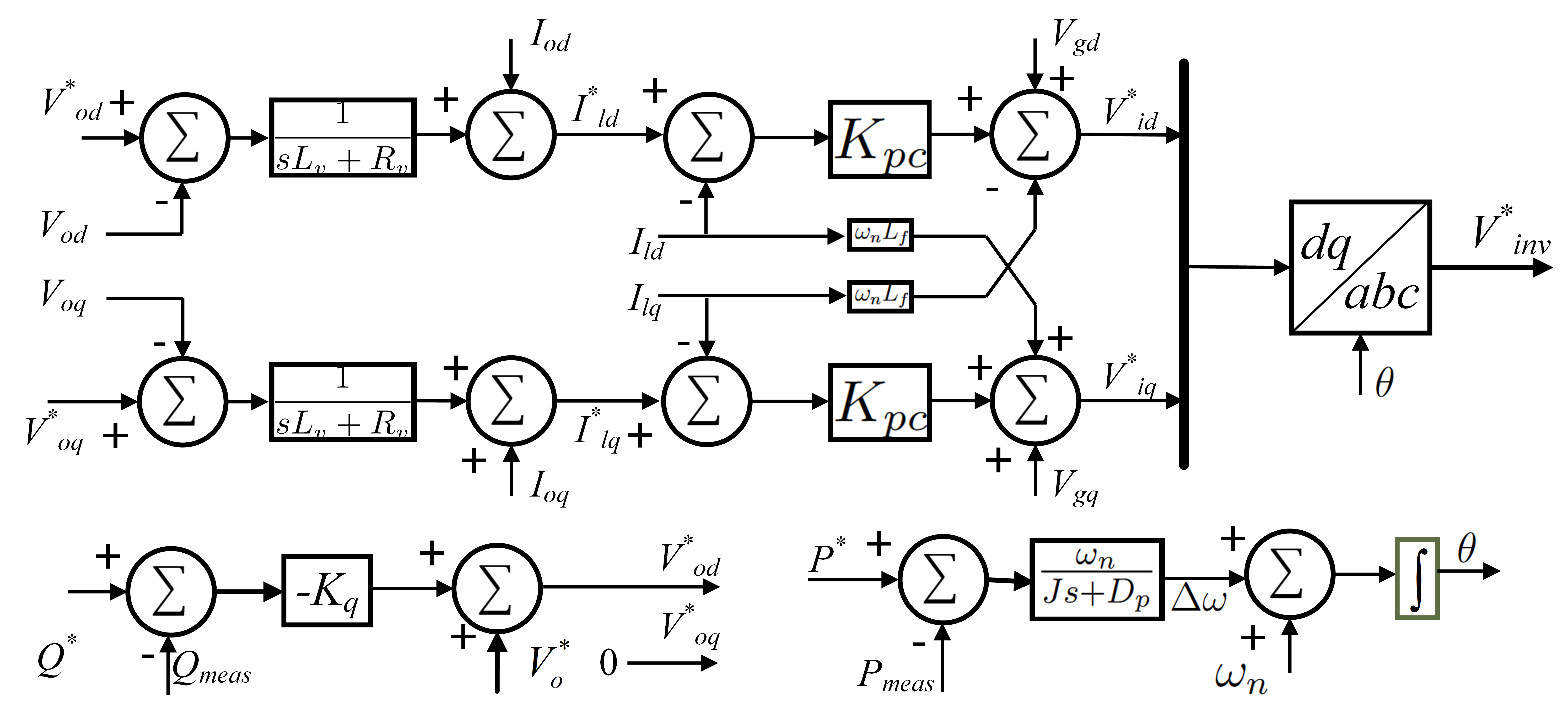}
         \caption{VAdm control diagram}
         \label{fig: VAdm Control Diagram}
     \end{subfigure}
    \caption{Control diagrams for VSM-based and VAdm-based GFM controllers employed in the WTGs. Source: \cite{ghimire2023gridforming}.}
    \label{fig: GFM Control Diagrams}
\end{figure}

The control block diagrams of the GFM control methods are presented in Fig. \ref{fig: GFM Control Diagrams}: Fig. \ref{fig: VSM Control Diagram} shows the virtual synchronous machine (VSM) based grid-forming control method and Fig. \ref{fig: VAdm Control Diagram} shows the virtual admittance (VAdm) based grid-forming control method. The presented GFM control methods are adapted from \cite{ghimire2023gridforming}.

The mathematical model of the outer power control loop of both VSM and VAdm are identical and can be written as:
\begin{eqnarray}
    \label{eq: VSM PQ loop 1}
    \theta &=& \int\left[\omega_n+\frac{1}{Js+D_p}(P^*-P_{meas})\right]dt,\\
    \label{eq: VSM PQ loop 2}
    V_{od}^* &=& V_n^* - K_q(Q^*-Q_{meas}),\quad V_{oq}^* = 0.
\end{eqnarray}
Here, $\theta$ is the phase defined by the converter control, $\omega_n:= 2\pi f$ is the base angular frequency of the power system with the frequency $f = 50$ Hz. $J$ and $D_p$ are the virtual inertia constant and damping constant respectivelpy, and $P^*$ and $P_{meas}$ are the active power references and measurements respectively. Similarly, $Q^*$ and $Q_{meas}$ denote the reactive power references and measurements respectively, $V_n^*$ denotes the voltage reference, and $V_{odq}^*$ denote the voltage references generated by the reactive power control loop.

The inner current and voltage control loops of VSM are PI-controller with dq-axis decoupling and are written as:
\begin{eqnarray}
    \label{eq: droop V loop 1}
    G_{PI}^{(v)}(s) &=& K_{pv} + \frac{K_{iv}}{s},\\
    \label{eq: droop V loop 2}
    I_{ldq}^* &=& I_{odq} + G_{PI}^{(v)}(s)(V_{odq}^*-V_{odq}) + j\omega_nC_fV_{odq},\\
    \label{eq: droop I loop 1}
    G_{PI}^{(c)}(s) &=& K_{pc} + \frac{K_{ic}}{s},\\
    \label{eq: droop I loop 2}
    V_{idq}^* &=& V_{gdq} + G_{PI}^{(c)}(s)(I_{odq}^*-I_{odq}) + j\omega_nL_fI_{ldq}.
\end{eqnarray}
Here, $K_{pv},\ K_{iv}$ and $K_{pc},\ K_{ic}$ are the respective PI-gains of the inner loop voltage and current controllers, and $V,\ I$ denote the measured voltage and currents while $V^*,\ I^*$ denote the reference voltage and currents respectively.

The inner loops for VAdm include a voltage correction via virtual admittance and proportional current controller and are written mathematically as:
\begin{eqnarray}
    \label{eq: VAdm Yvirt 1}
    Y_{virt}(s) &=& \frac{1}{s L_v + R_v},\\
    \label{eq: VAdm Yvirt 2}
    I_{ldq}^* &=& I_{odq} + (V_{odq}^* - V_{odq})Y_{virt}(s),\\
    V_{idq}^* &=& V_{gdq} + K_{P}^{(c)}(I_{ldq}^*-I_{ldq}) + j\omega_nL_fI_{ldq}.
\end{eqnarray}
Here, $L_v,\ R_v$ are the virtual inductance and resistance that define the virtual admittance $Y_{virt}(s).$


\section{WTGs with same GFM controls}\label{sec: Oscillation between WTGs with same GFM controls}
This section presents the observed oscillations in a WPP with WTG converters having same GFM control methods. Cases with matching control tunings and control parametric variations are presented.

\subsection{WTGs with same GFM controls and same control tunings}
\begin{figure}[htbp]
    \centering
    \begin{subfigure}[b]{0.49\textwidth}
         \centering
         \includegraphics[width=\textwidth]{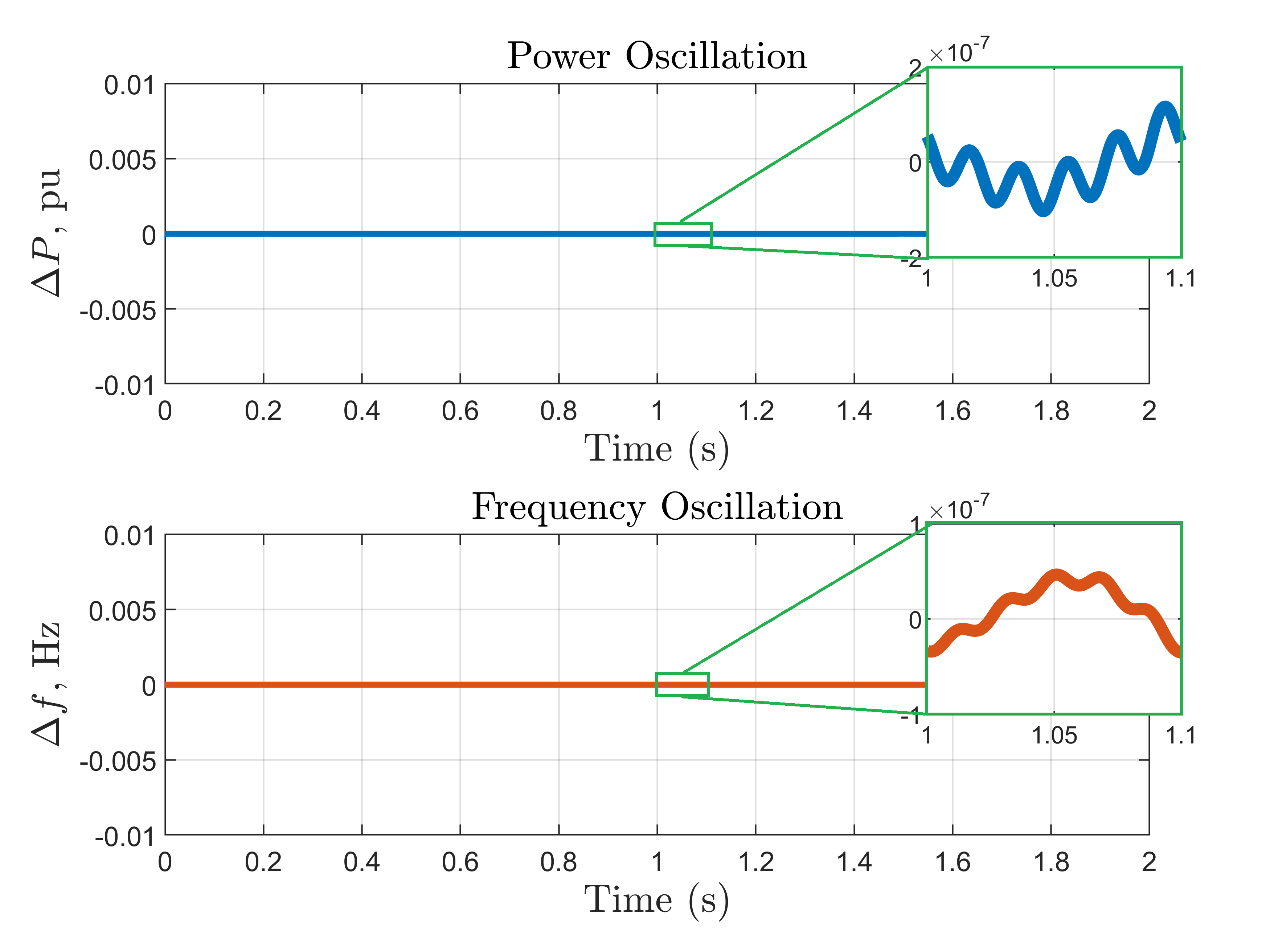}
         \caption{VSM vs VSM.}
         \label{fig: Case2_VSMVsVSM_SCRShift_Pf}
     \end{subfigure}
     \hfill
     \begin{subfigure}[b]{0.49\textwidth}
         \centering
         \includegraphics[width=\textwidth]{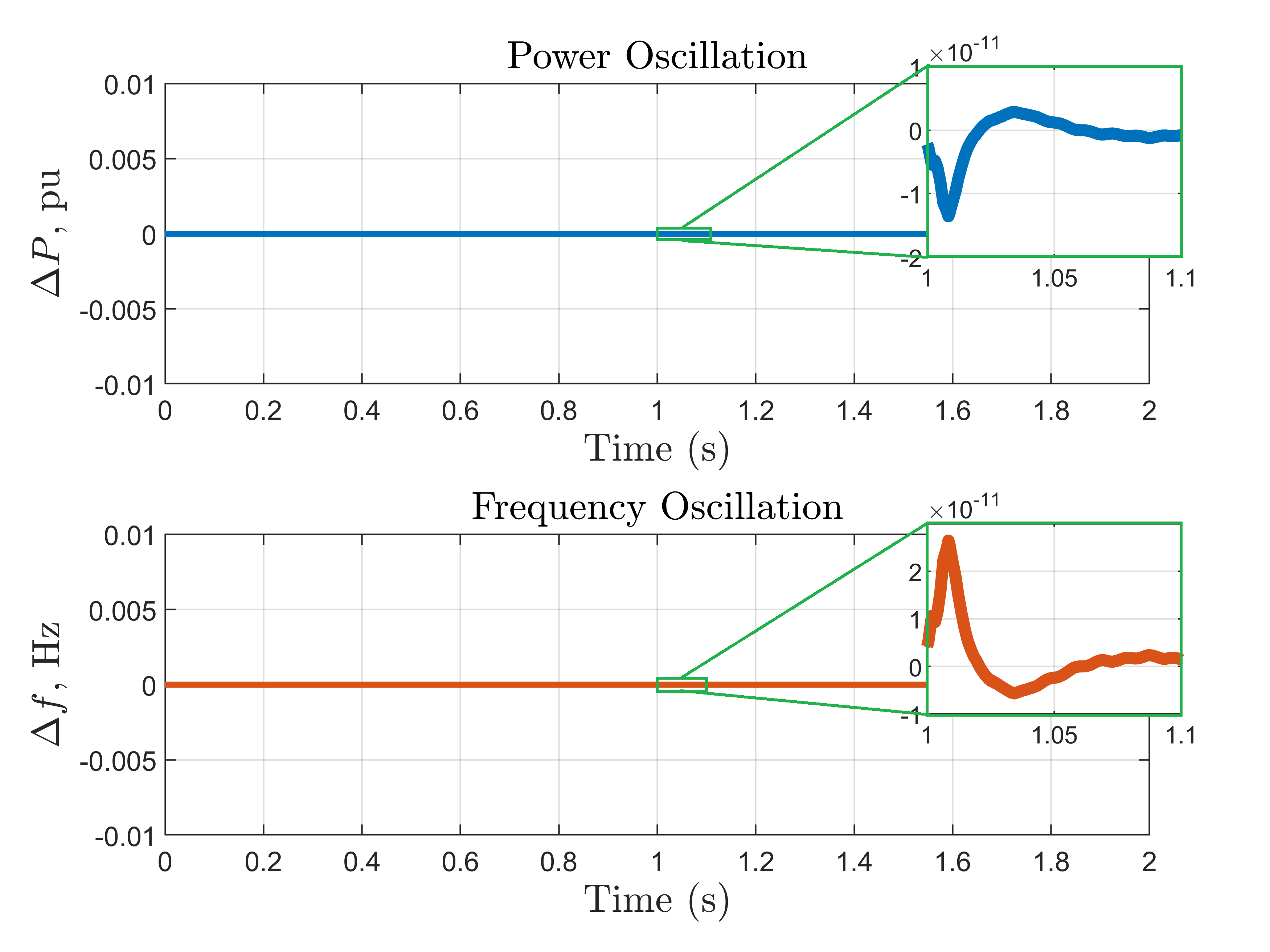}
         \caption{VAdm vs VAdm.}
         \label{fig: Case3_VAdmVsVAdm_SCRShift_Pf}
     \end{subfigure}
    \caption{Control oscillations in a system with same GFMs. A SCR change from 3.2 to 1.6 is applied at $t=1\ s.$}
    \label{fig: Case2 Case3 Same GFM}
\end{figure}

In this study case, both GFM-1 and GFM-2 in Fig. \ref{fig: TwoConv} are equipped with the same GFM control method with the same control tunings. The grid SCR is changed from 3.2 to 1.6 at t=1 s. The resulting power and frequency oscillations are presented in Fig. \ref{fig: Case2 Case3 Same GFM}. It was observed that there is no significant oscillation between the converters, either during steady state or during or following the SCR change event. As seen in Fig. \ref{fig: Case2_VSMVsVSM_SCRShift_Pf}, the oscillation magnitude for the case setup with two VSM controls is of the order of $10^{-7}$ and for the case setup with two VAdm controls (Fig. \ref{fig: Case3_VAdmVsVAdm_SCRShift_Pf}) oscillations are of the order of $10^{-11}$.

The results demonstrate that for converters with the same GFM control and matched control tunings, the power and frequency oscillations are negligible in magnitude. However, it must be noted that there are no measurement delays and latency in the control response included in the models used in this paper. Thus, the presented results do not include the effect of such delays on system oscillations.

\subsection{WPP with same GFM control and different inertia constant}
\begin{figure}[htbp]
    \centering
    \includegraphics[width=0.5\textwidth]{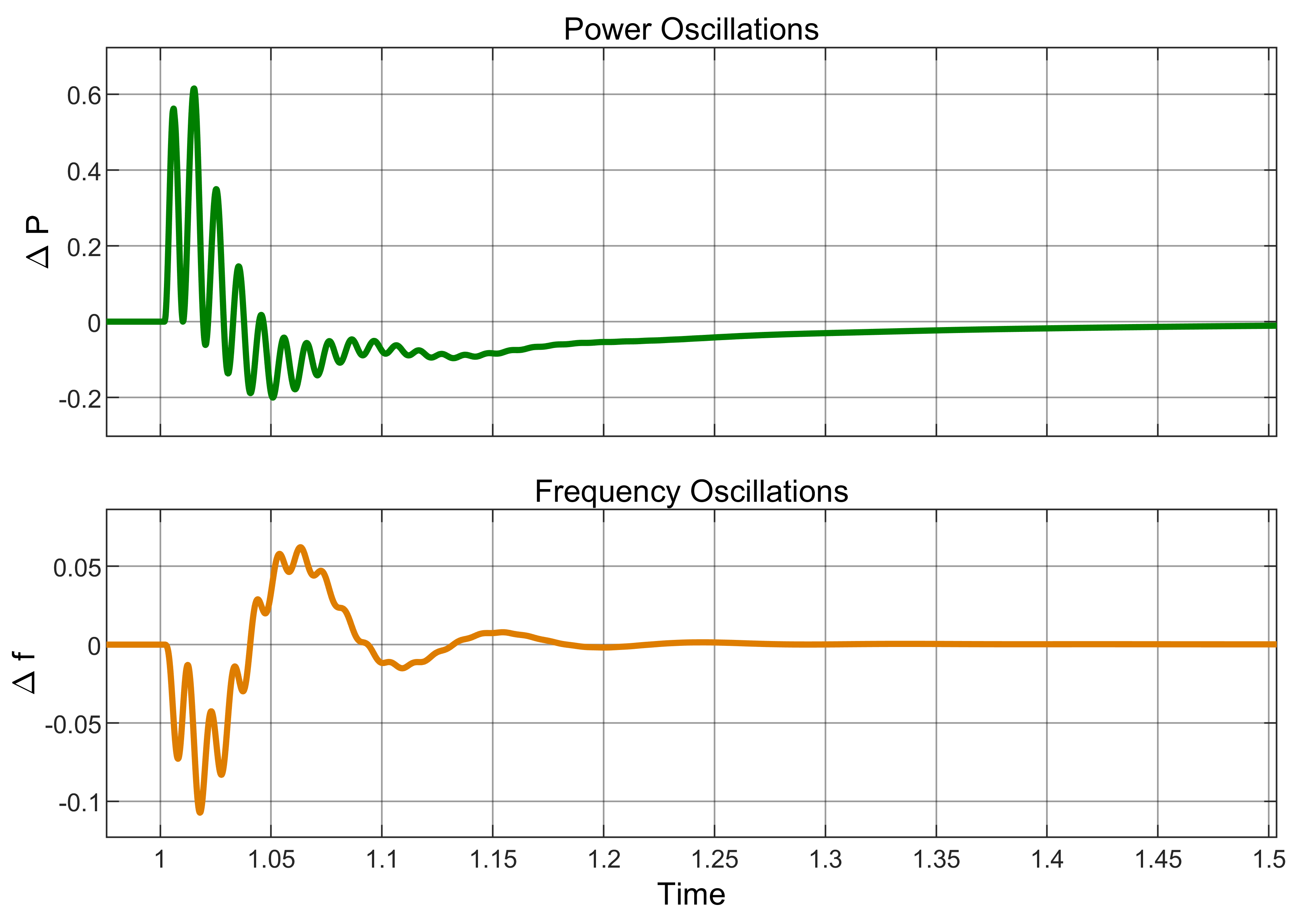}
    \caption{Oscillations between VSM and VSM0H controls following a grid SCR change from 3.2 to 1.6.}
    \label{fig: VSM VSM0H}
\end{figure}

This section studies the effect of different controller tunings on system oscillations. Both GFM-1 and GMF-2 in Fig. \ref{fig: TwoConv} are equipped with VSM control, however a parametric modification is performed on the the VSM control of GFM-1 WTG: its inertia component is removed such that $J = 0$. i.e. GFM-1 is now VSM0H or zero-inertia VSM control while GFM-2 is still the nominal VSM control. An SCR change from 3.2 to 1.6 is applied to the system at $t = 1\ s$. The results in Fig. \ref{fig: VSM VSM0H} show the transient response where $\approx$100 Hz power oscillations, and similar frequency oscillations are observed. However, there are no sustained oscillations once the transients clear out. The system gets fully re-synchronized 200-250 ms after the disturbance and becomes frequency-stable.

\section{WTGs with different GFM controls}\label{sec: Oscillation between WTGs with different GFM controls}
This section investigates the case of multiple GFM control methods in the same WPP. In the study cases presented herewith, GFM-1 and GFM-2 in the network topology in Fig. \ref{fig: TwoConv} are equipped with VSM and VAdm respectively. The different studies performed and the results obtained are presented herewith.

\subsubsection{Oscillation FFT during steady state and transients}
The system in Fig. \ref{fig: TwoConv} is subjected to a SCR change from 3.2 pu to 1.6 pu at time $t = 1$ s. Both the GFM converters shown in Fig. \ref{fig: GFM Control Diagrams} attempt to control the voltage and frequency at their terminals before the respective turbine transformers. During steady state operation and also during a transient, these converters attempt to control the voltage and frequency at the respective points. However, since the converters are not communicating and they're both controlling nodes with high electrical proximity, the power delivered by one is counterbalanced by the other. Further, due to the high dependency of the frequency set by the converters on their generated power, the frequency set by one converter is also counterbalanced by that set by the other leading to power and frequency oscillations between the converters.

\begin{figure}[htbp]
    \centering
    \begin{subfigure}[b]{0.24\textwidth}
         \centering
         \includegraphics[width=\textwidth]{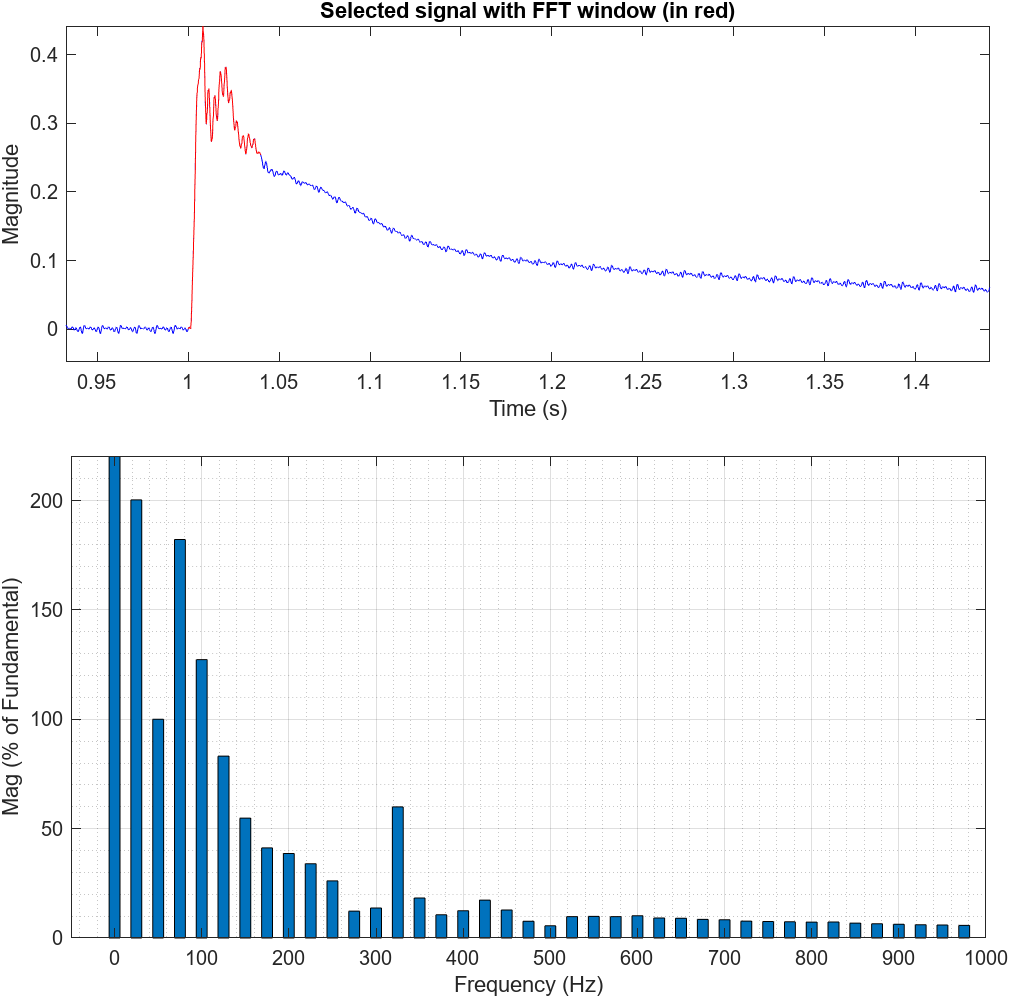}
         \caption{Power oscillations $P_1 - P_2$.}
         \label{fig: Pwr Osc FFT FltPt}
     \end{subfigure}
     \hfill
     \begin{subfigure}[b]{0.24\textwidth}
         \centering
         \includegraphics[width=\textwidth]{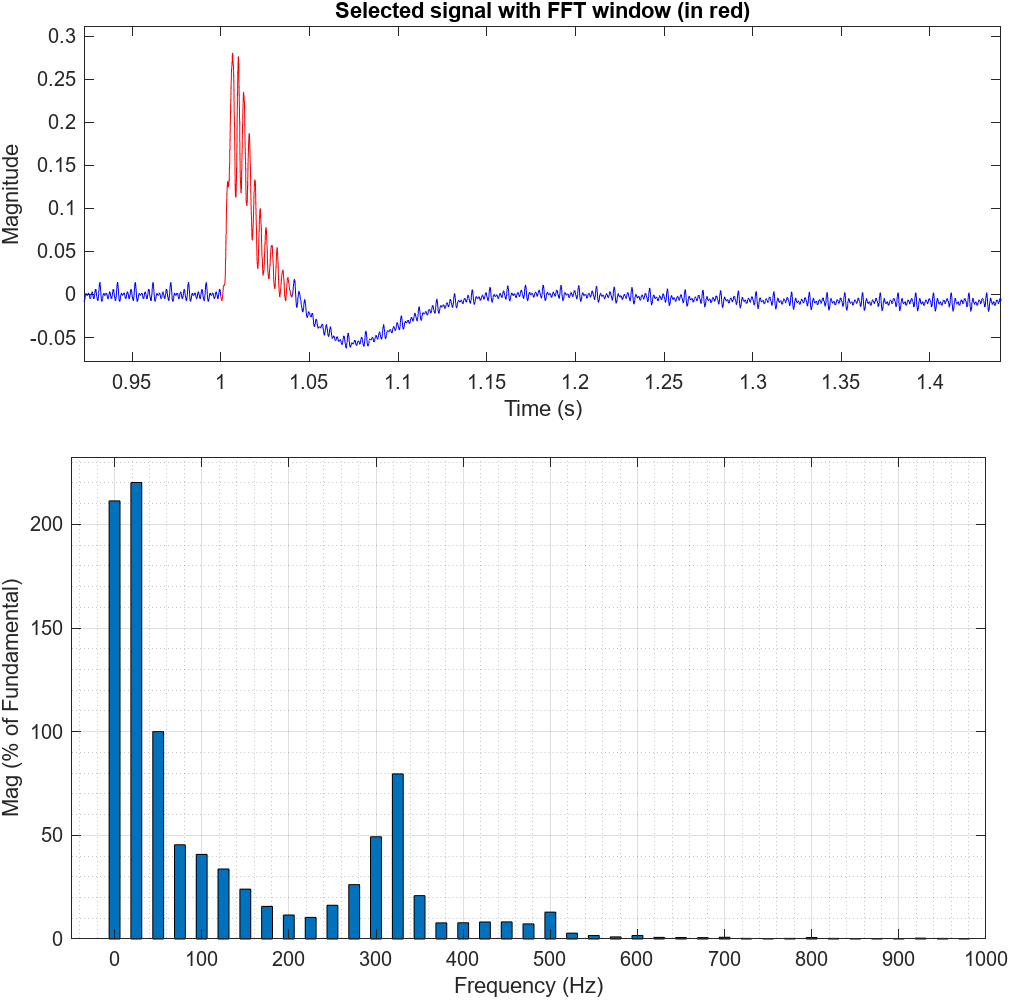}
         \caption{Frequency oscillations $f_1 - f_2$.}
         \label{fig: Freq Osc FFT FltPt}
     \end{subfigure}
    \caption{Fourier transform of the selected oscillation signals for two cycles after the SCR change event.}
    \label{fig: Pwr Freq Osc FFT Fault Point}
\end{figure}

\begin{figure}[htbp]
    \centering
    \begin{subfigure}[b]{0.24\textwidth}
         \centering
         \includegraphics[width=\textwidth]{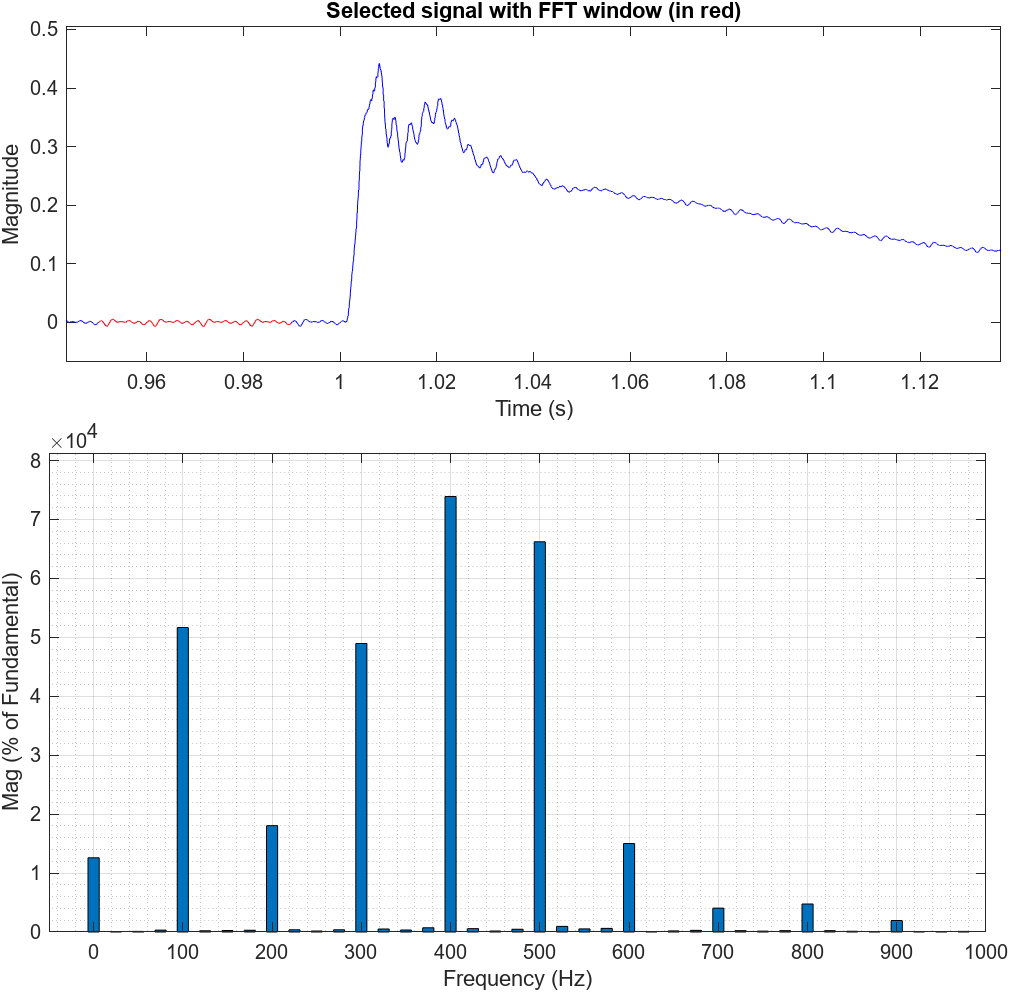}
         \caption{Power oscillations $P_1 - P_2$.}
         \label{fig: Pwr Osc FFT Steady State}
     \end{subfigure}
     \hfill
     \begin{subfigure}[b]{0.24\textwidth}
         \centering
         \includegraphics[width=\textwidth]{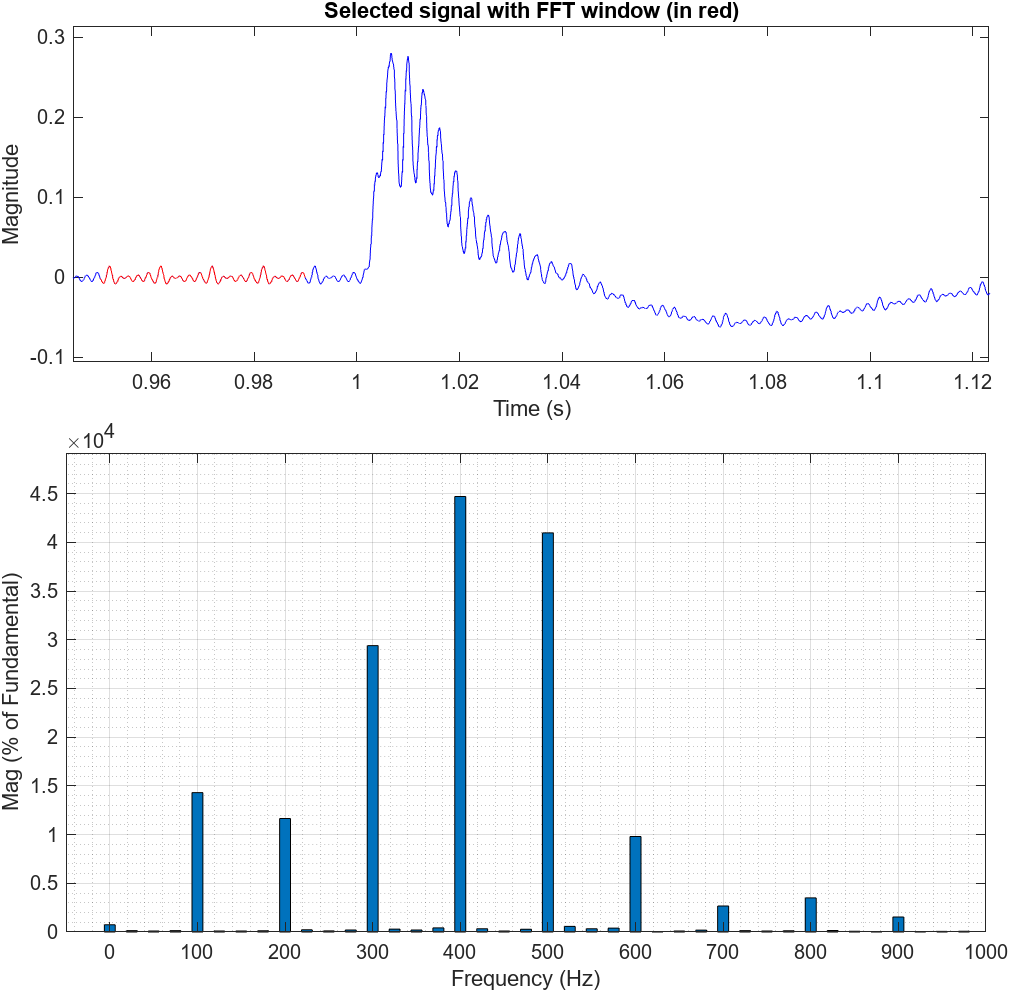}
         \caption{Frequency oscillations $f_1 - f_2$.}
         \label{fig: Freq Osc FFT Steady State}
     \end{subfigure}
    \caption{Fourier transform of the selected oscillation signals for two cycles during steady state.}
    \label{fig: Pwr Freq Osc FFT Steady State}
\end{figure}

The power and frequency oscillations shown in Fig. \ref{fig: Pwr Freq Osc FFT Fault Point} and Fig. \ref{fig: Pwr Freq Osc FFT Steady State} exhibit an amplitude of 0.01 pu and 0.01 Hz respectively. They also have some similar nature in their frequency content as the converter control defines the converter frequency based on power mismatch.

Fig. \ref{fig: Pwr Osc FFT FltPt} and Fig. \ref{fig: Freq Osc FFT FltPt} show that the sub-synchronous and near-synchronous oscillation frequencies are prevalent during the fault transient with 25 Hz component being dominant in both active power and frequency oscillations.

\subsubsection{Sensitivity of oscillations to electrical proximity between converters}
\begin{figure}[htbp]
    \centering
    \includegraphics[width = 0.49\textwidth]{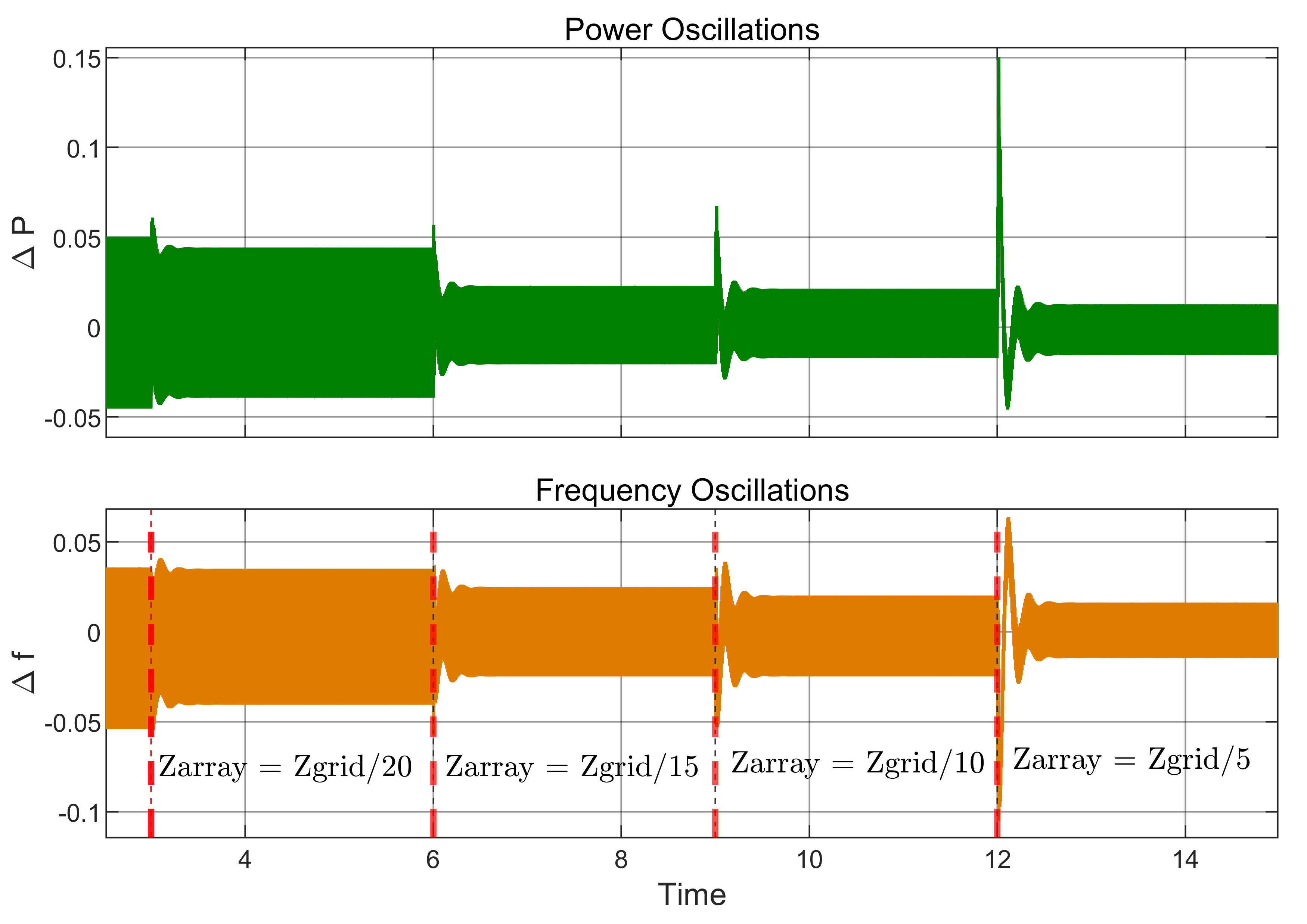}
    \caption{Inter-converter oscillations with change in electrical distance between the converters.}
    \label{fig: OscWithElectricalDistance}
\end{figure}

The electrical distance between the two converters in the presented system is $2(Z_{array}+Z_{tf})$ which includes the array cable impedance and transformer impedance of each of the power trains respectively. Now, keeping the transformer impedance $Z_{tf}$ constant, the array cable impedance $Z_{array}$ is varied from a minimum value $Z_{array} = Z_{grid}/20$ to its nominal value $Z_{array} = Z_{grid}/5$ to study the sensitivity of the oscillations to the electrical proximity between the two GFM WTGs. Fig. \ref{fig: OscWithElectricalDistance} presents the power and frequency oscillations between these GFM WTGs in each of these cases. It is evident from the figures that the control oscillations reduce significantly in amplitude with the reduction in electrical proximity between the WTGs. The reason is that the converter control systems are now controlling two distant nodes which facilitates more independent control without the need for communication, thus leading to less control interactions between the WTG converter control systems. Appropriate sizing of the array cables to facilitate the required power flow during all contingent cases without unnecessarily over-sizing it can alleviate the oscillation issues.

\subsubsection{Sensitivity of oscillations to grid SCR}
\begin{figure}[htbp]
    \centering
    \includegraphics[width = 0.49\textwidth]{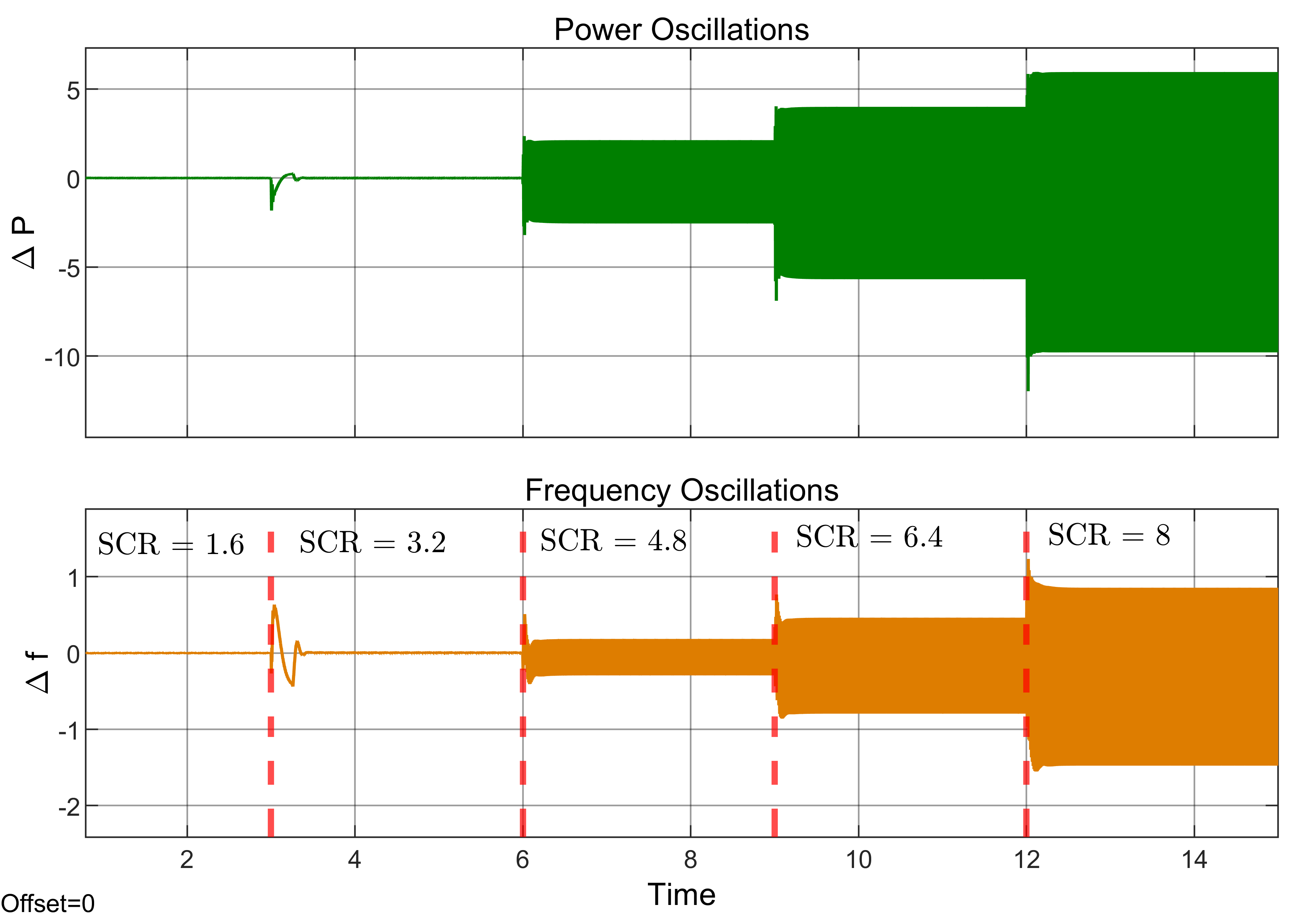}
    \caption{Inter-converter oscillations with change in Grid SCR.}
    \label{fig: OscWithGridSCR}
\end{figure}

Similarly, the grid SCR was increased from the nominal value of $SCR = 1.6$ to $SCR = 8$ keeping the electrical proximity between the GFM WTGs at their nominal value. The effect of five different SCR values on power and frequency oscillations is presented in Fig. \ref{fig: OscWithGridSCR}. It is evident that the control interactions depend heavily on grid SCR: as the grid SCR increases, the oscillations grow in magnitude. High grid SCR representing a strong grid suggests that the ideal voltage source of the grid (noting that the grid is modelled as a Thevenin-equivalent source) is in close electrical proximity to the controlled nodes of both converters. Now, although any disturbance in the system will mostly be counteracted by the grid, the GFM WTGs will still try to control their respective control nodes which are now in closer electrical proximity to the grid. Thus, the GFM WTGs now attempt to control a point close to the stiff-voltage source which is not possible, and instead end up oscillating against each other.

It must be noted that although the very high oscillations seen in cases with $SCR\ge 4.8$ in Fig. \ref{fig: OscWithGridSCR} are not seen in practical cases as the WTG protective controls like current limiting and FRTs become active. The results presented are based on WTG control models with disabled current limiting and FRT controls to facilitate the demonstration of the severity of the oscillations in a strong grid connected case.

\section{Potential Solution Methods}\label{sec: Potential Solution Methods}
Conventional inter-area and inter-machine oscillations in electrical power systems had a different nature than the inter-converter oscillations in modern power systems. As demonstrated here, the frequency components of the oscillations are dominantly super-synchronous and their dependence is mostly on system parameters and control systems rather than operating points. Since tools like power system stabilizers (PSS) and frequency washout filters -- which were quite useful to solve oscillations and instability in conventional power systems -- operate on low-frequency oscillations, they are unable to solve the oscillation issues observed in modern power systems and demonstrated herewith. Thus, methods like proper parameter tuning based on eigenvalue analyses and mode participation factors, and system design to maintain a safe electrical distance between different GFM-controlled WTG can aid in resolving these issues. Regarding the sensitivity of SCR on system stability, instead of modelling the grid as a stiff grid behind an impedance, an inertial model of the grid can be beneficial to properly understand and address the oscillation issues. Further, small-signal frequency domain models to pinpoint the states participating in any particular oscillation modes can help tuning the related control parameters and stabilize the system. Regarding the oscillations not severe in weak grids, but severe in the strong grids in Fig. \ref{fig: OscWithGridSCR}, different parameter type selection frameworks could be employed, however, this requires technological advancements to estimate grid SCR during operation. Designing and testing controls parameters which can withstand a wide range of grid SCR and electrical proximity of other sources could help select an appropriate control tuning which is robust to all those tests of grid SCR and electrical proximity with other sources.

\section{Conclusion}\label{sec: Conclusion}
Power and frequency oscillations between the same and different GFM converters were studied in an aggregated model of an offshore wind power plant. The results showed that the same GFM converters with matching tuning do not exhibit significant oscillations. The order of magnitude of the observed oscillations in such cases were infinitesimally small, thus negligible. However, parameter tunings can affect the system in such a way that they exhibit power and frequency oscillation. Further, systems with different GFM converters exhibited mild to severe levels of power and frequency oscillations depending on several network parameters. It was observed that the electrical proximity between the converters had a significant effect on oscillation amplitudes. Grid SCR had an even stronger effect on these oscillations: the converters exhibited higher amplitude oscillations in a strong-grid connected case. Thus appropriate parametric tuning and matching control and parameter selection is recommended for WTGs in the same WPP or for WPPs in close electrical proximity.

\small
\section*{Legal Disclaimer}%
Figures and values presented in this paper should not be used to judge the performance of Siemens Gamesa (SGRE) technology as they are solely presented for demonstration purposes. Any opinions or analyses contained in this paper are the opinions of the authors and are not necessarily the same as those of SGRE.

\section*{Acknowledgement}
The authors would like to acknowledge the support of Siemens Gamesa Renewable Energy (SGRE), Technical University of Denmark (DTU), and Innovation Fund Denmark. The work of S.G. was supported by Innovation Fund Denmark under the project Ref. no. 0153-00256B.

\bibliography{bib/Bibliography}
\bibliographystyle{ieeetr}

\end{document}